\documentclass[a4paper,11pt]{scrartcl}
\usepackage[centertags]{amsmath}
\usepackage{amsfonts}
\usepackage{amssymb}
\usepackage{amsthm}
\usepackage{nicefrac}
\usepackage{titlesec}
\usepackage{pifont}
\usepackage[ansinew]{inputenc}
\usepackage{graphicx}
\usepackage{color}

\DeclareMathOperator{\DR}{DR}
\DeclareMathOperator{\CDR}{CDR}

\begin{document}

\title{How to assess case-finding in chronic diseases: Comparison of different indices.}

\author{Ralph Brinks\footnote{ralph.brinks@ddz.uni-duesseldorf.de}\\
Institute for Biometry and Epidemiology\\German Diabetes Center\\Düsseldorf, Germany}

\date{}

\maketitle

\begin{abstract}
Recently, we have proposed a new illness-death model that comprises
a state of undiagnosed chronic disease preceding the diagnosed disease.
Based on this model, the question arises how case-finding can be assessed
in the presence of mortality from all these states.
We simulate two scenarios of different performance of case-finding and 
apply several indices to assess case-finding in both scenarios.
One of the prevalence based indices leads to wrong conclusions. Some
indices are partly insensitive to distinguish the quality of case-finding.
The incidence based indices perform well. If possible, incidence based 
indices should be preferred.
\end{abstract}


\section{Introduction} 
Many chronic diseases have a preclinical phase, when the disease is principally detectable but has 
not been diagnosed yet. Examples are cardiovascular disease, diabetes, chronic obstructive 
pulmonary disease, depression, or dementia. 

We call all collective activities and efforts, in which cases of a specific chronic disease not 
known to the health services are searched for, \emph{case-finding}. Case-finding in this sense is 
a very broad term, which comprises screening, application of diagnostic tests and all other 
activities and policies of detecting cases of a chronic disease.
Efforts in case-finding vary substantially, both globally and temporally. Geographical variations 
are due to different health systems, available resources and differences in the risk for certain 
diseases in a region or country. For example, efforts in case-finding for dementia are likely 
to be higher in older 
populations, e.g., in the industrialised countries. Reasons for temporal changes in case-finding of 
chronic diseases are manifold as well. Besides technical progress in making 
diagnostic tests cheaper and more practicable, varying awareness of patients and physicians may 
lead to an earlier or later diagnosis of the disease.

Early detection of diseases might be important for two reasons. First, in many cases it is 
favourable if the disease is treated early to make effective treatment possible, e.g. cancer 
\cite{Smi04}, diabetes \cite{Mel10}, or chronic kidney disease \cite{Wha10}. Second, patients with 
an undiagnosed disease already have an elevated risk for unfavourable outcomes. For example, persons 
with undiagnosed diabetes have an about 50\% increased risk of all-cause mortality compared to a 
healthy person \cite[Table 2]{Gor15}. Another example is undiagnosed chronic obstructive pulmonary 
disease, which is rather frequent \cite{Bas10} and associated with a severe loss of quality of 
life \cite{Mir09}.

Given the enormous importance of case-finding in chronic diseases, we want to examine different 
measures of how to quantitatively assess the activities of case-finding of a specific chronic 
disease. The question arises what epidemiological measures are suitable for describing the 
performance of case-finding on the population level.

\bigskip

We give an example from the epidemiology of diabetes. Table \ref{tab:KORA} shows the prevalence 
undiagnosed and diagnosed diabetes for men
and women in two age groups of the KORA study \cite{Rat03}. In men, we see that with 
increasing age the percentages of undiagnosed and diagnosed diabetes increase. 
The situation is different in women. As the age increases,
the prevalence of undiagnosed and diagnosed diabetes rises and lowers, respectively.
The key question of this article is:  which situation
is more desirable with respect to case-finding, the situation of men or the one of women?

\begin{table}[ht!]
\centering
\caption{Sex-specific prevalence of undiagnosed and diagnosed diabetes
in two age-groups of the KORA study \cite{Rat03}.}
\begin{tabular}{c|cc|cc}
Age      & \multicolumn{2}{|c|}{Undiagnosed diabetes (in \%)} & \multicolumn{2}{|c}{Diagnosed diabetes (in \%)} \\
(years)  & Men  & Women & Men  & Women  \\ \hline
60--64   &  8.1 &  7.3 &  7.2 &    9.7  \\
65--69   &  8.9 &  8.2 & 13.3 &    8.2  \\
\end{tabular}
\label{tab:KORA}
\end{table}

Before we try to answer this question, we describe the underlying epidemiological model
for assessing the performance of case-finding.

\section{Epidemiological model}
\begin{figure}[ht]
\centering
\includegraphics[width=120mm,keepaspectratio]{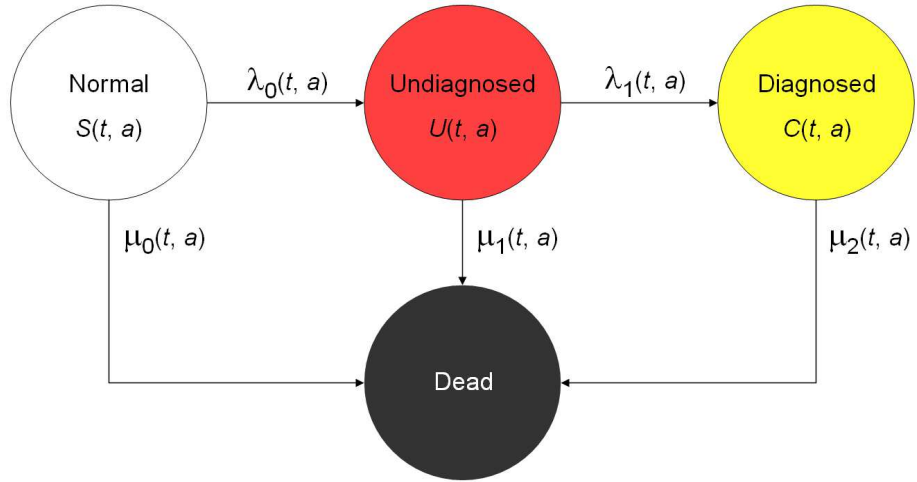}
\caption{Chronic disease model. Persons in the state \emph{Normal}
are healthy with respect to the disease under consideration. After
onset of the disease, they change to state \emph{Undiagnosed} and 
later to the state \emph{Diagnosed}. The absorbing state \emph{Dead} can be
reached from all other states. The percentages of persons in the states and
the transition rates depend on calendar time $t$ and age $a.$}
\label{fig:TirionModel}
\end{figure}

If we are interested in evaluating the efforts of case-finding in a population,
we consider each person being exactly in one of the depicted states shown in Figure \ref{fig:TirionModel}:
\emph{Normal}, i.e., healthy with respect to the considered chronic disease, \emph{Undiagnosed}, 
\emph{Diagnosed} or \emph{Dead}. At the birth each person is in the \emph{Normal}
state (here we just consider diseases acquired after birth). During the life course,
the person may contract the disease and enters the \emph{Undiagnosed} state.
In a screening program, or as the disease becomes symptomatic, or
during some routine medical examinations, the disease may be diagnosed
and the person enters the \emph{Diagnosed} state. Persons may decease from
any of these three states. Note that we are just considering chronic diseases. 
Hence, backward steps from the \emph{Undiagnosed}
state to the \emph{Normal} state are not possible. Similarly, we
assume that once a person has got a diagnosis the disease remains
detected for ever. Thus, we assume that there is no loss of such information.

In \cite{BriBMC} we described 
relations between the transition rates $\lambda_\ell(t, a), ~\ell = 0, 1,$ and $\mu_k(t, a), ~k=0, 1, 2$, in 
the model and the percentages of persons in the states. As in \cite{BriBMC} let $p_k(t, a)$ 
denote the fraction of persons aged $a$ at time $t$ in state $k, ~k=0, 1, 2.$ For example,
$p_1(t, a)$ is the fraction of persons in the population who are aged $a$ at time $t$ and are in the 
\emph{Undiagnosed} state ($k=1$).

\section{Assessment of case-finding}
In this section, we describe measures for assessing the performance of case-finding.
We distinguish figures based on the prevalences and figures based on the transition
rates between the states in the epidemiological model (cf. Figure \ref{fig:TirionModel}).

\subsection{Figures based on the prevalence}
The first figure to approach case-finding, is the proportion of detected cases from
the total cases \cite{Gre04}, i.e., 
\begin{equation}\label{e:wp1}
\wp_1 = \tfrac{p_2}{p_1+p_2}.
\end{equation}
Similar to the dark figure in criminology, the reciprocal of $\wp_1$ 
describes the factor the diagnosed cases have to be multiplied with to obtain 
the number of all cases of the chronic disease. 
Obviously, it holds $0 \le \wp_1 \le 1.$ A high value in $\wp_1$ is usually
interpreted as advantageous \cite{Gre04}.

\bigskip

Analogously, it may be useful to consider the ratio 
\begin{equation}\label{e:wp2}
\wp_2 = \tfrac{p_1}{p_0+p_1}.
\end{equation}

This ratio relates the number of persons in the undiagnosed state to all persons who do 
not have a diagnosis, i.e., the healthy and the undiagnosed. 
The idea behind the measure $\wp_2$ is that case-finding can be thought of 
distinguishing persons from a pool consisting of healthy and undiagnosed persons. This
pool of healthy and undiagnosed persons may be seen as the \emph{search space}. The search 
space is subject to the activities of case-finding. Once an undiagnosed person is 
undoubtedly identified
as a case, this person gets a diagnosis and is removed from the search space henceforth. 
As the disease under consideration is chronic, there is no way back into the search space.
In contrast to $\wp_1$, the figure $\wp_2$ just refers to the persons who are at risk
for a possible diagnosis. The fraction of persons with a diagnosis does not play a role.

Again, it holds $0 \le \wp_2 \le 1.$ Ideally, $\wp_2$ is 0, i.e., all undetected
cases are removed from the search space. The closer 
$\wp_2$ approaches 1, the more the search space is dominated by the undiagnosed persons.
Thus, a lower value of $\wp_2$ is advantageous in assessing case-finding.

\bigskip

If we calculate the measures $\wp_1$ and $\wp_2$ for the data in Table \ref{tab:KORA},
we obtain the values shown in Table \ref{tab:W-KORA}.

\begin{table}[ht!]
\centering
\caption{Age- and sex-specific ratios $\wp_1$ and $\wp_2$ for the data in Table \ref{tab:KORA}.}
\begin{tabular}{c|cc|cc}
Age      & \multicolumn{2}{|c|}{$\wp_1$ (in \%)} & \multicolumn{2}{|c}{$\wp_2$ (in \%)} \\
(years)  & Men  & Women & Men  & Women  \\ \hline
60--64   &  47  &  57   &    9 &   8    \\
65--69   &  60  &  50   &   10 &   9    \\
\end{tabular}
\label{tab:W-KORA}
\end{table}

For men, the measure $\wp_1$ increases from 47\% to 60\%
as the age increases. Thus, $\wp_1$ indicates that with increasing age the performance of
case-finding improves. However, $\wp_2$ indicates that the performance 
of case-finding in men deteriorates from the lower age class to the higher. Thus, the measures
$\wp_1$ and $\wp_2$ yield \emph{contradicting} findings in assessing the performance
of case-finding.

As the age increases, the measure $\wp_1$ for women decreases from 57\% to 50\%, 
which indicates a worsening of case-finding. Similarly, $\wp_2$ shows a worsening. 
In women, the figures $\wp_1$ and $\wp_2$ allow the \emph{same} conclusion. This
example from the KORA study shows that at least one of the figures $\wp_1$ or $\wp_2$
is not suitable for assessing the performance of case-finding. We will come back later
to this point.

\subsection{Figures based on the transition rates}
Apart from the figures based on the prevalences, we may consider figures based on the
transitions in the model. 
\subsubsection{Incidence rate ratio}
In \cite{BriBMC} we used the rate ratio 
$\DR = \tfrac{\lambda_1}{\lambda_0},$ which relates the 
instantaneous risk (hazard) of transiting to the diagnosed state to the risk of becoming an 
undetected case. As it is unlikely to be diagnosed immediately after entering the undiagnosed 
state, we introduce a delay parameter $\gamma, ~\gamma \ge 0,$ and define

\begin{equation*}
\DR_\gamma(t, a) = \frac{\lambda_1(t + \gamma, a + \gamma)}{\lambda_0(t, a)}.
\end{equation*}

Obviously, it holds $\DR = \DR_0.$

\bigskip

\subsubsection{Deaths without a diagnosis}
An important figure is the fraction of healthy persons aged $a$ at time $t$ who
become incident undiagnosed cases at time $t$ and die within $\gamma > 0$
time units without a diagnosis. As these persons do not have a diagnosis,
they never were treated. To develop this figure we first calculate the 
probability of dying $P^{(\text{dead})}_\gamma$ during the first $\gamma > 0$ 
time units in the undiagnosed state:
\begin{equation}\label{e:dead}
P^{(\text{dead})}_\gamma  (t, a) = \int_0^\gamma \mu_1    (t + s, a + s) \, 
                            \exp \Bigl ( - \int_0^s (\lambda_1 + \mu_1)(t + \tau, a + \tau) \mathrm{d}\tau \Bigr ) \, \mathrm{d}s.
\end{equation}

Then, the probability $P^{(\text{dead})}_\gamma$ is combined with the incidence rate $\lambda_0:$

\begin{equation}\label{e:C}
\Phi_\gamma(t,a) = \lambda_0(t,a) \, P^{(\text{dead})}_\gamma (t,a).
\end{equation}
Then, $\Phi_\gamma(t, a)$ is the number of death cases
per one healthy person aged $a$ at time $t$, who becomes an incident undiagnosed case 
at $t$ and dies within $\gamma$ time units without diagnosis. To get
an integer number, one may multiply with a power of 10, say, 100,000. 
These originally 100,000 healthy persons never had the chance of obtaining a treatment.

\subsection{Other figures}
In the field of infectious disease epidemiology, 
sometimes the \emph{case detection rate} ($\CDR$) is considered. The CDR is the
notification rate of incident cases over the total incidence rate. Roughly speaking, it is the 
proportion of detected \emph{incident} cases from the total \emph{incident} cases 
\cite{Bor04}. In our terminology, the $\CDR$ can be calculated as:
\begin{equation}\label{e:CDR}
\CDR = \DR_0 \, \wp_2.
\end{equation}
A proof for this relation can be found in the appendix.

\section{Simulation study} 
In the introducing example from the KORA study, we have seen that the prevalence-based 
measures $\wp_1$ and $\wp_2$ have come to contradicting conclusions about
the performance of case-finding in the male population. So far, it remains open which
measure is more appropriate for the assessment of case-finding.

In this section, we conduct a simulation study with two different settings to answer
this question. In one 
setting, the underlying (true) performance of case-finding worsens over time, 
whereas in the other setting
the true performance betters. For both settings, we calculate the introduced measures of assessing the 
case-finding by comparing two points in time. For setting up the simulation, we use
a system of partial differential equations with known transition rates.

\subsection{Partial differential equations}
Based on the epidemiological model in Figure \ref{fig:TirionModel},
we have shown that in a population without migration and with sufficiently smooth
transition rates $\lambda_\ell, ~\ell = 0, 1, ~\mu_k, k=0, 1, 2,$ the prevalences 
$p_k, ~k=0, 1, 2,$ are governed by a set of partial differential equations \cite{BriBMC}:

\begin{align}
(\partial_t + \partial_a) p_1 &= \lambda_0 - \bigl ( \lambda_0 + \lambda_1 + \mu_1 - \mu_0 \bigr ) \, p_1 - \lambda_0 \, p_2 + \lambda_0 \label{e:pde1}\\
(\partial_t + \partial_a) p_2 &= \lambda_1 \, p_1 - \bigl ( \mu_2 - \mu \bigr ) \, p_2 \label{e:pde2}.
\end{align}

The notation $\partial_x$ means the partial derivative with respect to $x, ~x \in \{t, a\}$. In Equations
\eqref{e:pde1} -- \eqref{e:pde2}, the term $\mu$ is the overall mortality (general mortality), which
can be written as
\begin{equation}\label{e:genMort}
\mu = p_0 \, \mu_0 + p_1 \, \mu_1 + p_2 \, \mu_2.
\end{equation}

The prevalence $p_0$ can be calculated by using the equation $p_0 = 1 - p_1 - p_2.$ 
Thus, together with the initial conditions $p_1(t , 0) = p_2(t , 0) = 0$ for all $t,$ the
system \eqref{e:pde1} -- \eqref{e:pde2} completely describes the dynamics of the disease 
in the considered population. 

\bigskip

For later use, we remark that Equations \eqref{e:pde1} -- \eqref{e:pde2} can be 
transformed into following system:
\begin{align}
(\partial_t + \partial_a) p_1 &= \lambda_0 \, ( 1 - p_2) - ( \lambda_0+\lambda_1 + \mu_1 - \mu_0 ) \, p_1 
                                   + p_1 \, z \label{e:pde3}\\
(\partial_t + \partial_a) p_2 &= - ( \mu_2 - \mu_0 ) \, p_2 + \lambda_1 \, p_1 + p_2 \, z \label{e:pde4},
\end{align}
where $z = p_1 \, (\mu_1-\mu_0) + p_2 \, (\mu_2-\mu_0).$

We will integrate the system \eqref{e:pde3} -- \eqref{e:pde4} using the 
\emph{Method of Characteristics} \cite{Pol01} and Runge-Kutta
integration \cite{Dah74}. All calculations are done with the software \textsl{R}
(The R Foundation of Statistical Computing).

\subsubsection{Transition rates between the states}
The incidence rate $\lambda_0$ is chosen to be 
\begin{equation}\label{e:l0}
\lambda_0(t, a) = \frac{1}{3} \times 10^{-4} \times (a-30)_+ \times 1.015^t,
\end{equation}
which in magnitude coarsely mimics the incidence of type 2 diabetes in males \cite{Car08}.

\bigskip

The rate $\lambda_1$ is assumed to be a sigmoid function with a steep increase between
the 20th and 50th year of age:
\begin{equation}\label{e:l1}
\lambda_{1}(t, a, \varepsilon) = \frac{0.02}{1+\exp\bigl(-0.25 \times (a-35) \bigr)}  \times \bigl ( 1.015 + \varepsilon \bigr )^t.
\end{equation}
This rate mimics an hypothetical awareness for type 2 diabetes, which is assumed to 
increase after an onset at $a=20.$

\medskip

From Eq. \eqref{e:l0}, we observe that the rate $\lambda_0$ 
rises with calendar time $t$. 
For the rate $\lambda_1$ we choose two 
simulation settings (A) and (B) with $\varepsilon_A = -0.016$ and 
$\varepsilon_B = 0.01.$ In settings A and B, the annual change 
$(1.015 + \varepsilon)$ of $\lambda_1$
is negative and positive, respectively. This
will lead to an accumulation of undetected cases over calendar time in setting A and a slowly 
decreasing reservoir of undetected cases in setting B.

\begin{figure}[ht]
\centering
\includegraphics[width=150mm,keepaspectratio]{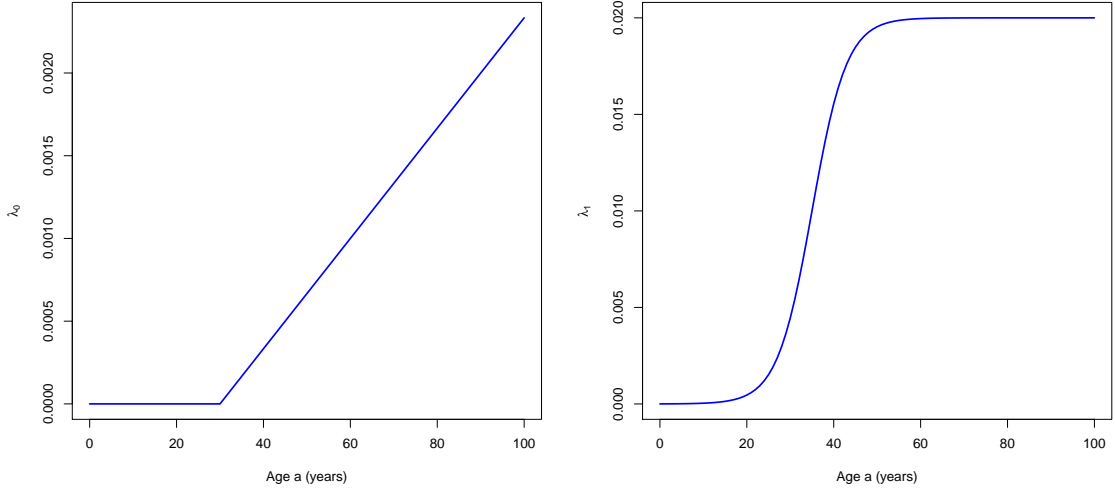}
\caption{Age courses of $\lambda_0$ (left) and $\lambda_1$ (right) in year $t = 0.$}
\label{fig:Lambdas}
\end{figure}

Figure \ref{fig:Lambdas} shows the age courses of $\lambda_0$ and $\lambda_1$ for $t=0.$ Note
that for $t=0$ the rates $\lambda_1$ in setting A and B are the same: 
$$\lambda_1(0, a, \varepsilon_A) = \lambda_1(0, a, \varepsilon_B), ~\text{for all } a.$$

We choose Gompertz mortality rates, \cite[Eq. (9.1)]{Pre00}: 
\begin{equation}\label{e:SM-mort}
\mu_k(t, a) = \exp(-\xi_k + \eta_k \, a ) \times (1 - \rho_k)^t, ~k=0, 1, 2,
\end{equation}
with the coefficients as in Table \ref{tab:Coeff}. The mortality rates of many 
populations approximately follow a Gompertz law, and the numbers roughly reflect
the mortality of German males in the past century. Calendar time $t$
in this sense is the time (in years) after 1900.

For further references 
and a critical discussion of the Gompertz law of mortality see, 
for instance, \cite{Fin12}. The yearly decrements
$\rho_k$ are chosen with a view to trends in mortality. The choice 
$\rho_2 > \rho_1$ is motivated by the fact that medical progress is
reaching treated persons more than untreated.

\begin{table}[ht!]
\centering
\caption{Coefficients of the Gompertz mortality rates $\mu_k, ~k=0, 1, 2$.}
\begin{tabular}{c|ccc}
$k$ &  $\xi_k$ & $\eta_k$ & $\rho_k$ \\ \hline
$0$ &    9.8   &   0.09   &  0.015   \\
$1$ &    9.7   &   0.10   &  0.015   \\
$2$ &    9.2   &   0.11   &  0.030   \\
\end{tabular}
\label{tab:Coeff}
\end{table}

\subsubsection{Prevalence}
Integrating Eq. \eqref{e:pde3} -- \eqref{e:pde4} with the rates given in Eq. \eqref{e:l0} 
-- \eqref{e:l1} and the Gompertz mortalities (Eq. \eqref{e:SM-mort} and 
Table \ref{tab:Coeff}), we obtain $p_1$ and $p_2$ for the years $t= 100$ and $t=110$
in both simulations as shown in Table \ref{tab:Prev}.

\begin{table}[ht!]
\centering
\caption{Age-specific prevalence of undiagnosed ($p_1$) and diagnosed ($p_2$) disease in the simulation.}
\begin{tabular}{|c||cc|cc||cc|cc|} \hline
Age     & \multicolumn{4}{|c||}{Setting A}                 & \multicolumn{4}{|c|}{Setting B} \\
(years) & \multicolumn{2}{|c|}{$p_1$ (in \%)} & \multicolumn{2}{|c||}{$p_2$ (in \%)} & 
          \multicolumn{2}{|c|}{$p_1$ (in \%)} & \multicolumn{2}{|c|}{$p_2$ (in \%)}          \\
        & $t=100$ & $t=110$ & $t=100$ & $t=110$ & $t=100$ & $t=110$ & $t=100$ & $t=110$      \\ \hline
   45   &   1.4   &   1.7   &  0.10   &  0.12 &   0.74  &   0.73  &  0.79  &  1.0            \\
   60   &   4.7   &   5.5   &  0.81   &  0.94 &   1.6   &   1.5   &  4.0   &  5.0            \\
   75   &   8.4   &   9.8   &  2.2    &  2.6  &   2.3   &   2.1   &  8.5   & 10              \\
   90   &   9.3   &  11     &  2.7    &  3.2  &   2.5   &   2.3   &  9.8   & 12              \\ \hline
\end{tabular}
\label{tab:Prev}
\end{table}

\begin{figure}[ht]
\centering
\includegraphics[width=150mm,keepaspectratio]{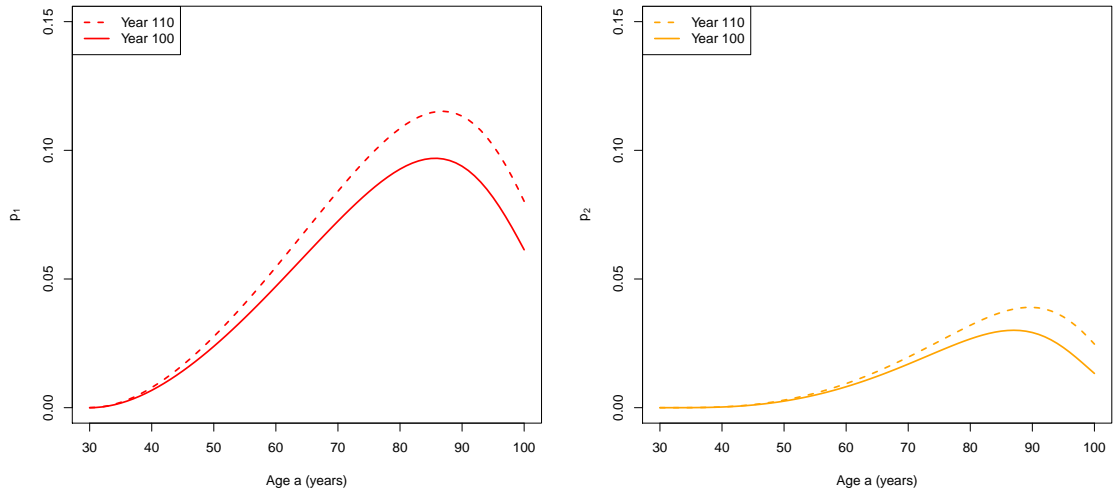}
\caption{Age courses of $p_1$ (left) and $p_2$ (right) in the years $t=100$ 
(solid line) and $t=110$ (dashed lines) in simulation setting A.}
\label{fig:PrevA}
\end{figure}

\begin{figure}[ht]
\centering
\includegraphics[width=150mm,keepaspectratio]{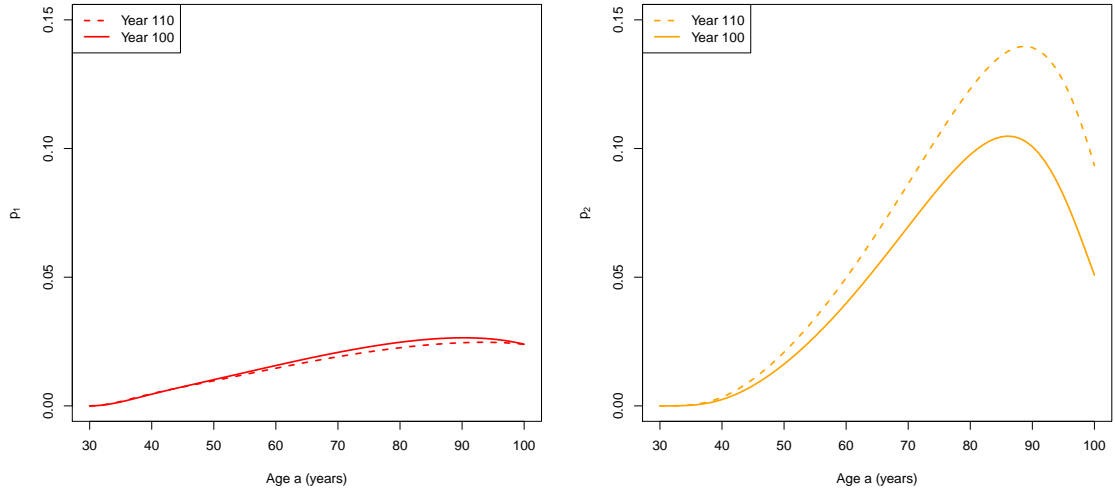}
\caption{Age courses of $p_1$ (left) and $p_2$ (right) in the years $t=100$ 
(solid line) and $t=110$ (dashed lines) in simulation setting B.}
\label{fig:PrevB}
\end{figure}

The age courses of the prevalence of the undiagnosed
($p_1$) and the diagnosed disease ($p_2$) in the years $t_1=100$ and $t_2=110$ 
are depicted in Figures \ref{fig:PrevA} and \ref{fig:PrevB} for both simulation settings. 
The age courses are realistic for a widespread chronic disease like diabetes or 
certain types of cancer. 
 
In both simulation settings, A and B, we see that the prevalence of the diagnosed chronic
disease is increasing from year 100 to 110 in nearly all age classes. 
The prevalence of the undiagnosed disease is also growing
for virtually all ages in simulation A. However in setting B, the prevalence of the undiagnosed
is remaining constant or slightly decreasing -- despite $\lambda_0$ increases 
over time for each age group. As expected, we observe an accumulation of undetected cases in 
setting A and a slowly decreasing reservoir of undetected cases in setting B during the 
period from year 100 to year 110. By comparing the prevalences of 
the undiagnosed and diagnosed disease in settings A and B, we can conclude that the
performance of case-finding worsens in setting A and improves in setting B.

\subsection{Assessing the case-finding}

\subsubsection{Prevalence based figures}
\begin{figure}[ht]
\centering
\includegraphics[width=150mm,keepaspectratio]{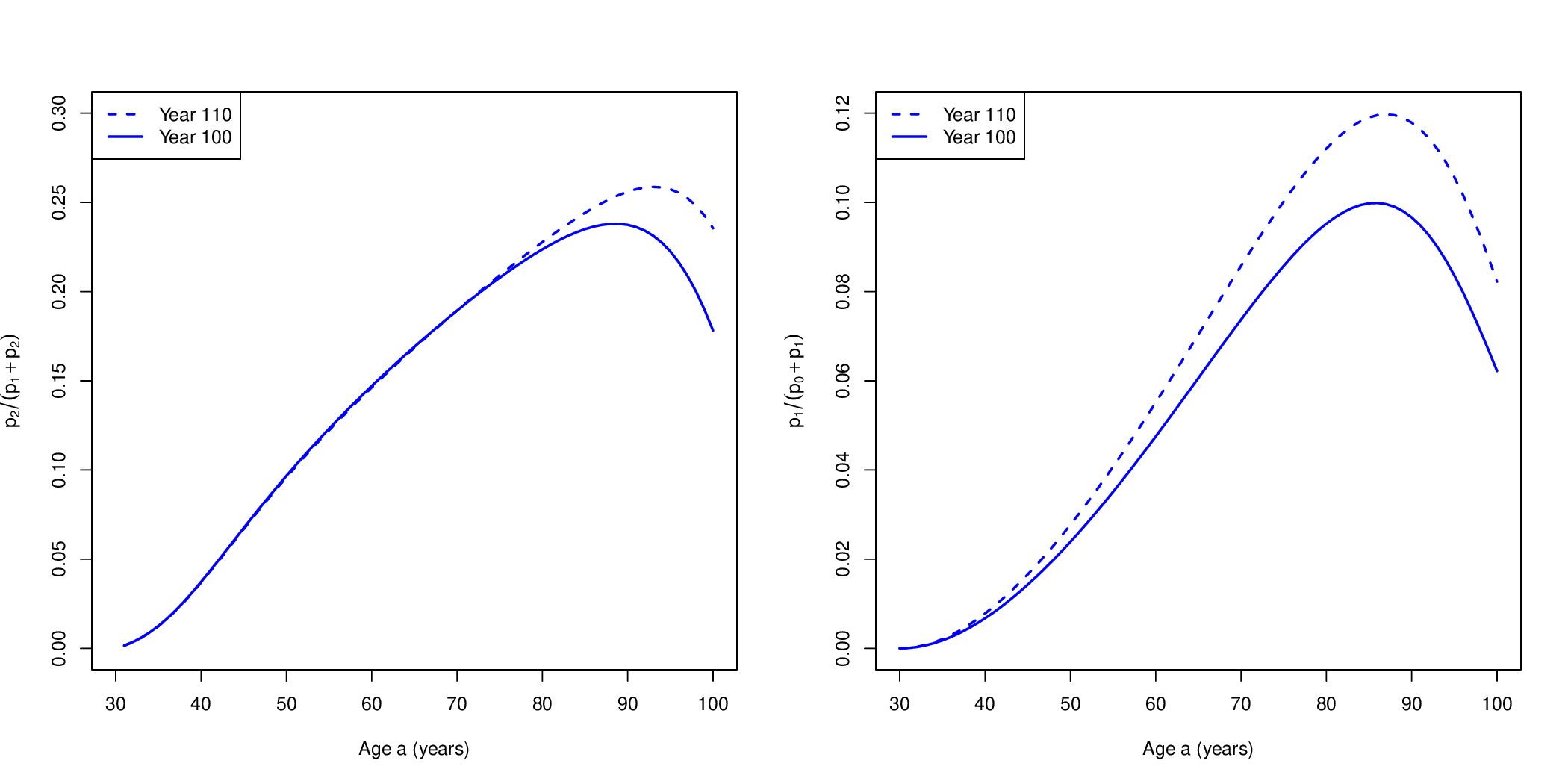}
\caption{Age courses of the case-finding measures $\wp_1$ (left) and $\wp_2$ (right) 
in the years $t = 100$ (solid line) and $t=110$ (dashed line) in simulation A.}
\label{fig:PRA}
\end{figure}

\begin{figure}[ht]
\centering
\includegraphics[width=150mm,keepaspectratio]{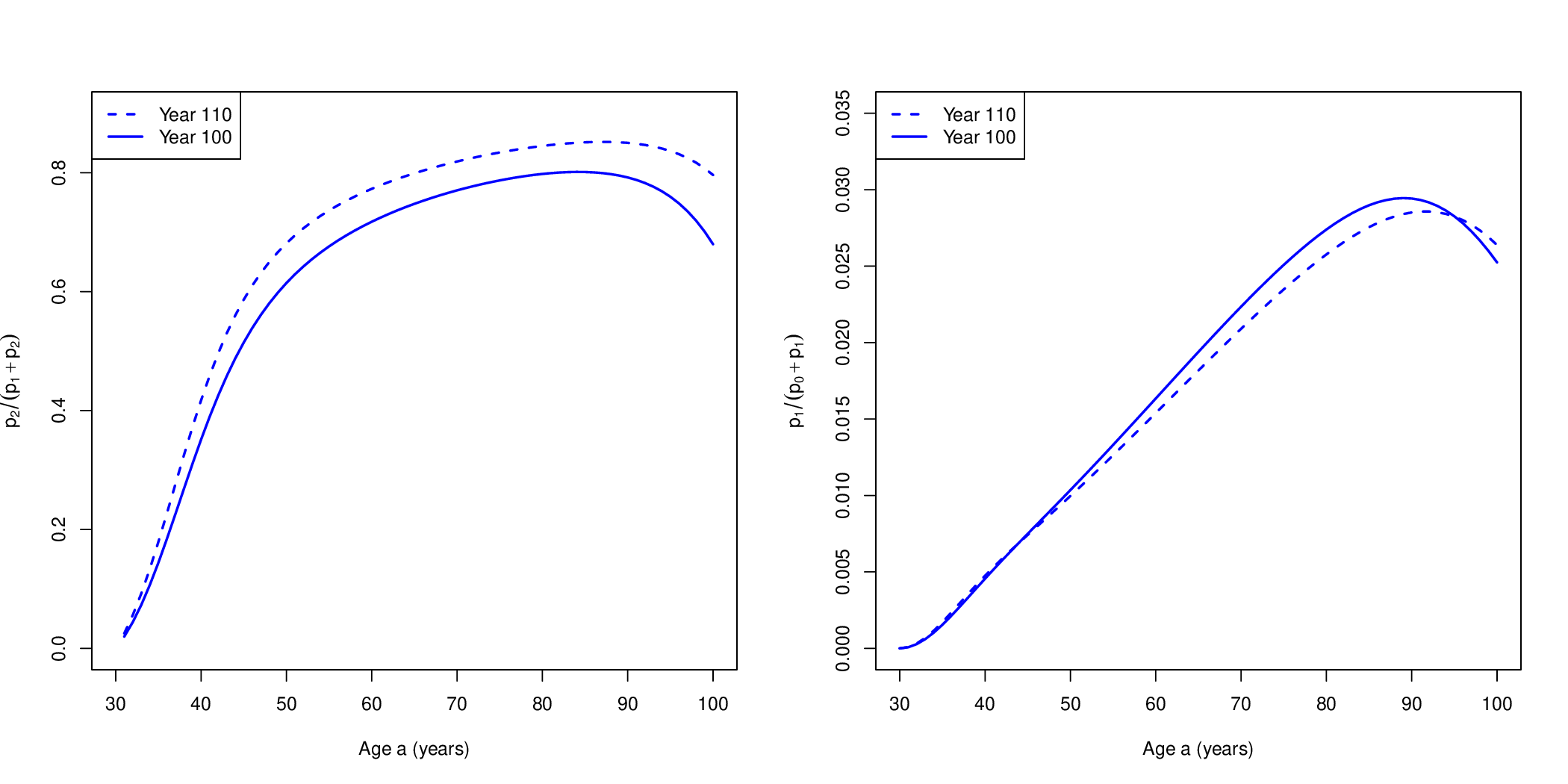}
\caption{Age courses of the case-finding measures $\wp_1$ (left) and $\wp_2$ (right) 
in the years $t = 100$ (solid line) and $t=110$ (dashed line) in simulation B.}
\label{fig:PRB}
\end{figure}

If we calculate the ratios $\wp_1$ and $\wp_2$ for some ages
in years 100 and 110 in both simulation settings, we obtain the results as presented in 
Table \ref{tab:P12}. Graphical presentations are given in Figures \ref{fig:PRA} and 
\ref{fig:PRB}.

\begin{table}[ht!]
\centering
\caption{Age-specific detection ratios $\wp_1$ and $\wp_2$
in both simulation settings.}
\begin{tabular}{|c||cc|cc||cc|cc|} \hline
Age     & \multicolumn{4}{|c||}{Setting A}                 & \multicolumn{4}{|c|}{Setting B} \\
(years) & \multicolumn{2}{|c|}{$\wp_1$ (in \%)} & \multicolumn{2}{|c||}{$\wp_2$ (in \%)} & 
          \multicolumn{2}{|c|}{$\wp_1$ (in \%)} & \multicolumn{2}{|c| }{$\wp_2$ (in \%)}     \\
        & $t=100$ & $t=110$ & $t=100$ & $t=110$ & $t=100$ & $t=110$ & $t=100$ & $t=110$      \\ \hline
   45   &    6.7  &   6.7   &   1.4   &    1.6  &    51   &     59  &   0.75  & 0.74         \\
   60   &   15    &  15     &   4.8   &    5.5  &    72   &     77  &   1.6   & 1.5          \\
   75   &   21    &  21     &   8.6   &   10    &    79   &     83  &   2.5   & 2.3          \\
   90   &   22    &  23     &   9.5   &   11    &    80   &     84  &   2.7   & 2.6          \\ \hline
\end{tabular}
\label{tab:P12}
\end{table}

In simulation setting A, we see from Table \ref{tab:P12} and Figure \ref{fig:PRA} 
that for all ages the ratio $\wp_1$ remains the same or rises between $t = 100$ and $t=110.$
For example, in the age group of 75 year-old persons, the ratio $\wp_1$ 
is 21\% for both years. Thus, based on
$\wp_1$ one would judge that the situation of case-finding remains the same
and slightly improves for the higher age groups during 
the period 100 -- 110. However if we consider $\wp_2$ in setting A, 
we get the indication that the situation worsens. For all age groups, $\wp_2$ increases 
during 100 to 110, which indicates that 
during that period the percentage of undiagnosed persons in the search space increases.
Hence, the measures $\wp_1$ and $\wp_2$ yield contradicting findings about
the performance of case-finding in the years $t=100$ and $t=110$ in simulation setting A.

In simulation setting B, Table \ref{tab:P12} and Figure \ref{fig:PRB} show that
$\wp_1$ indicates an improvement during the period from year 100 to year 110. 
$\wp_2$ also reveals an improvement for the ages 40 to 95. 

To sum up, we can say that the figure $\wp_2$ assesses both settings A and B correctly, 
whereas $\wp_1$ does not value setting A as being negative with respect to case-finding..

\clearpage

\subsubsection{Figures based on transitions between the states}
\begin{figure}[ht!]
\centering
\includegraphics[width=150mm,keepaspectratio]{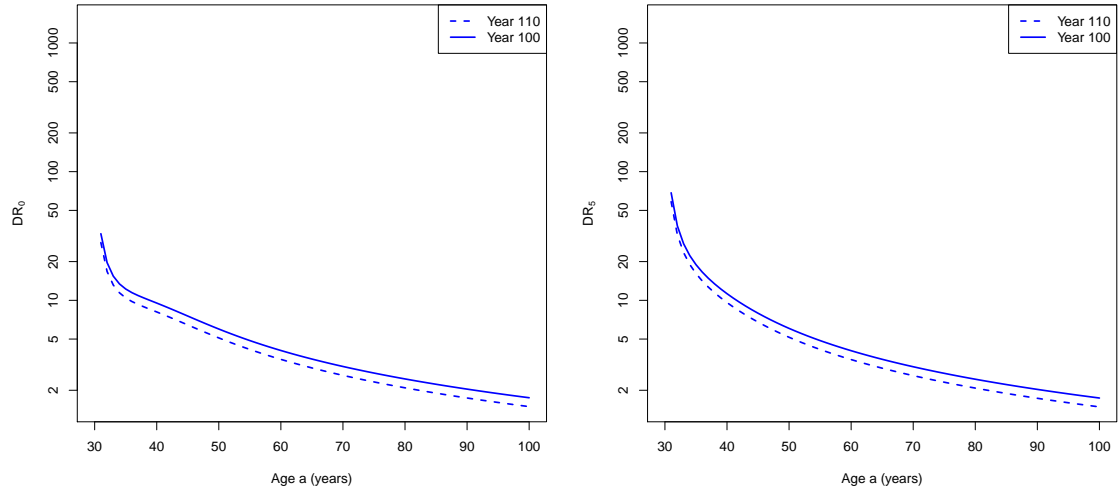}
\caption{Age courses of the detection ratios $\DR_0$ (left) and $\DR_5$ (right) 
in the years $t = 100$ (solid line) and $t=110$ (dashed line) in simulation setting A.}
\label{fig:DRA}
\end{figure}

\begin{figure}[ht!]
\centering
\includegraphics[width=150mm,keepaspectratio]{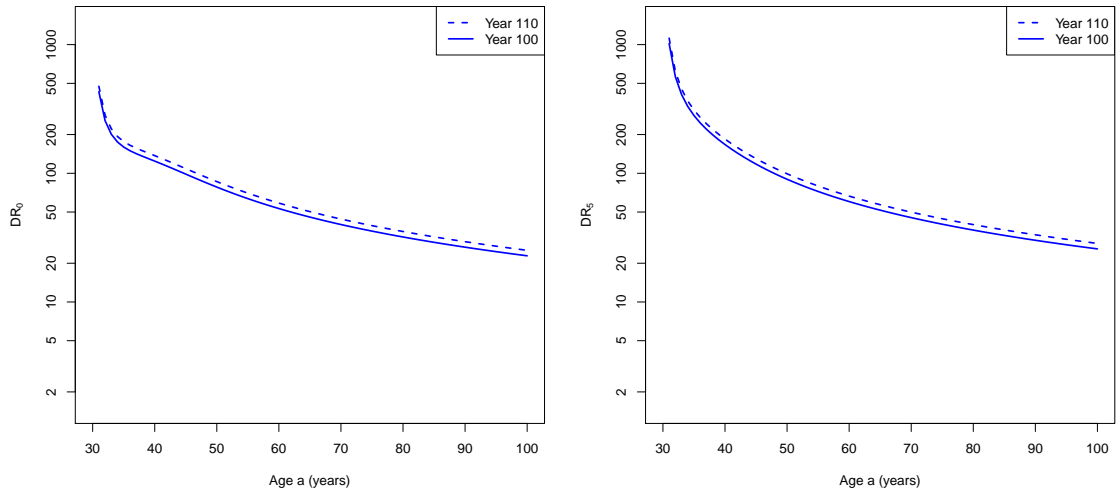}
\caption{Age courses of the detection ratios $\DR_0$ (left) and $\DR_5$ (right) 
in the years $t = 100$ (solid line) and $t=110$ (dashed line) in simulation setting B.}
\label{fig:DRB}
\end{figure}

Figures \ref{fig:DRA} and \ref{fig:DRB} show the age-specific detection rates $\DR_0$ and $\DR_5.$
Although different in magnitude, both measures $\DR_0$ and $\DR_5$ indicate
correctly that the case-finding performance worsens from 100 to 110 in setting A and
improves in setting B.

\begin{figure}[ht!]
\centering
\includegraphics[width=150mm,keepaspectratio]{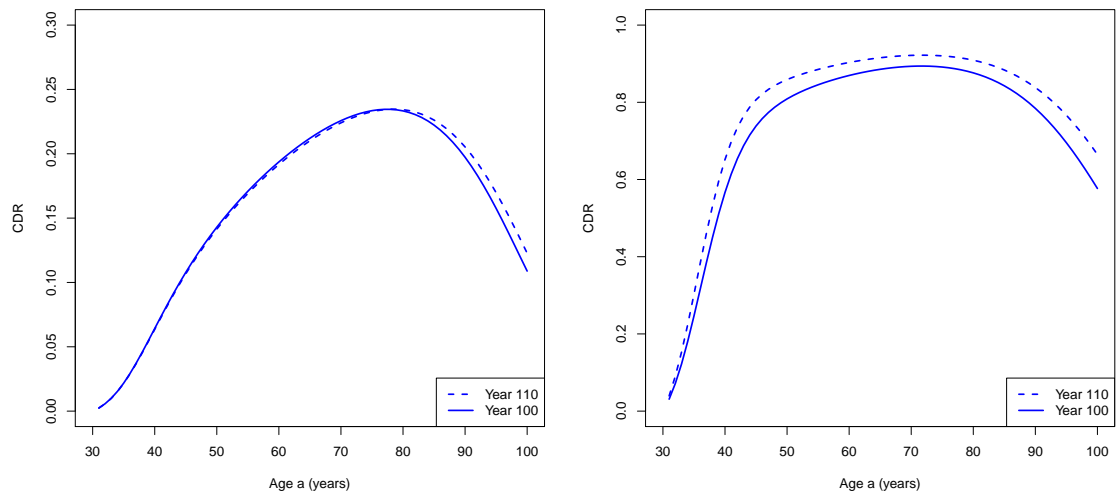}
\caption{Age courses of the case detection ratio $\CDR$ in simulation setting A (left) 
and setting B (right) in the years $t = 100$ (solid line) and $t=110$ (dashed line) in simulation setting B.}
\label{fig:CDR}
\end{figure}

In Figure \ref{fig:CDR} we see the age courses of the $\CDR$ in the different simulation
settings A (left) and B (right). We see that in setting A the $\CDR$ fails to
indicate a lower performing case-finding for a wide range of ages. In setting B, the 
$\CDR$ correctly values the situation as an improvement of case-finding.

\bigskip

The results of assessing case-finding in the simulation settings A and B by the figure $\Phi_5$  
are shown in Table \ref{tab:Cases}. Let us consider, for example, the age group 90. 
In year $t=100$, out of 100,000 healthy persons 886 persons contract the undiagnosed
disease. Ten years later, this number increases to 1029. Both numbers just depend 
on $\lambda_0$. Thus, they are valid for both simulation settings. 
Out of these 886 and 1029 incident undiagnosed cases in years 100 and 110, within five
years after onset of the disease 420 and 437, respectively, die without a diagnosis in setting A.
Thus, we have an increase of persons who never had the chance for a treatment, which
clearly indicates a worsening of case-finding performance.
In setting B, the number of these fatalities is considerably lower. In the years
100 and 110, we observe 259 and 235 death cases without diagnoses, respectively, a
considerably lower number.
Hence, the measure $\Phi_5$ indicates that the situation worsens in setting A and improves in setting B.

\begin{table}[ht!]
\centering
\caption{Incident cases and fatalities $\Phi_5$ from the undiagnosed state during 5 years (per 100,000 healthy persons).}
\begin{tabular}{|c||cc|cc|cc|} \hline
  Age   & \multicolumn{2}{|c|}{Incident cases (undiagnosed)}  & \multicolumn{4}{|c|}{100,000 $\Phi_5$} \\ 
(years) & \multicolumn{2}{|c|}{(per 100,000 healthy persons)} & \multicolumn{2}{|c}{Setting A} & \multicolumn{2}{|c|}{Setting B} \\ 
        & $t = 100$ & $t = 110$ & $t = 100$ & $t = 110$ & $t = 100$ & $t = 110$ \\ \hline
  45    &  222 &  257 & 1.600  & 1.600 &  0.951 &   0.828 \\ 
  60    &  443 &  514 & 14.11 &  14.12 & 8.202 &   7.109 \\ 
  75    &  665 &  772 & 89.59 &  90.35 & 52.62 &  46.02 \\ 
  90    &  886 & 1029 & 420.0 & 437.2 & 259.0 & 235.1 \\ \hline
\end{tabular}
\label{tab:Cases}
\end{table}

\clearpage

Table \ref{tab:SumMeasures} sums up the findings of assessing the performance of case-finding in the
different simulations settings. The measures $\wp_1$ and $\CDR$ at least partly fail
to indicate the deterioration of case-finding in setting A, whereas the figures
$\wp_2$, $\DR_0$, $\DR_5$ and $\Phi_5$ assess both settings correctly.

\begin{table}[ht!]
\centering
\caption{Summary of the different measures of assessing case-finding. A plus (+) or a 
minus (--) sign denotes whether the measure indicates an improvement or a deterioration 
of case-finding over time.}
\begin{tabular}{|c||c|c|} \hline
Measure               & Simulation A & Simulation B \\ \hline
$\wp_1$               & +/--  & +   \\
$\wp_2$               & --    & +   \\
$\DR_0$               & --    & +   \\
$\DR_5$               & --    & +   \\
$\CDR$                & +/--  & +   \\
$\Phi_5$              & --    & +   \\ \hline
\end{tabular}
\label{tab:SumMeasures}
\end{table}

\clearpage

\section{Discussion} 
Based on a system of partial differential equations
we set up a simulation study with an temporally increasing (setting A) and decreasing (setting B)
quality of case-finding in an hypothetical chronic disease. Then, we applied different figures 
to assess and compare the performance of case-finding at two points in time. 
Some of the measures were not able to judge the differences in 
both simulation settings correctly. Table \ref{tab:SumMeasures} shows how the different
settings have been assessed by the different figures.

We found that the measures $\wp_1$ and $\CDR$ are unsuitable measures 
to assess the case-finding performance in chronic diseases, because the unfavourable 
situation of an temporally increasing reservoir of undiagnosed cases (setting A) has not
valued as being negative.

The figure $\wp_2$ correctly values settings A and B. Thus, it
is sensitive measure for the improvement and degradation of case-finding. Similarly,
the detection ratios $\DR_\gamma$ correctly assess the different simulation settings
for $\gamma = 0$ and $\gamma = 5.$ The figure $\Phi_5$ is an important measure, which 
refers to a cohort of healthy persons
who contract the disease but never get the chance of being treated.


\bigskip

There are other figures to assess case-finding. For example, the mean sojourn 
time (MST) in the preclinical phase may be considered, see \cite{Uhr10} for a review. 
Usually, a low MST is considered advantageous. However, the MST may be low if the
mortality from the undiagnosed state is high. Thus, the MST is not an appropriate
figure for evaluating case-finding.

\bigskip

Compared to the other figures, the measure 
$\wp_2 = \tfrac{p_1}{p_0+p_1}$ has the advantage that it just require prevalence 
data, which can be obtained from cross-sectional studies. Those measures including the incidence 
rates either require costly follow-up data or the application of specialized estimation techniques
\cite{BriBMC}.

\bigskip

In this work, we applied these measures to data about undiagnosed and diagnosed prevalence of diabetes 
from the KORA study.
The ratio $\wp_2$ unveils that in men and women, the performance of case-finding is better
in the age group 60--64 compared to the the age group 65--69. The reasons for this may
be individual or societal, but a detailed analysis is beyond the scope of this article.

\section*{Appendix}
The $\CDR$ is the proportion of incident cases being diagnosed \cite{Bor04}, i.e.,
the notification rate of incident cases over the total incidence rate.
Let $\lambda'$ be the notification rate, which is the number of
detected cases transiting from the search space to the \emph{Diagnosed} state per unit time. 
Thus, the denominator is the number of persons in the combined state of \emph{Normal} and 
\emph{Undiagnosed}. In the model in Figure \ref{fig:TirionModel} the rate $\lambda_1$ refers
to transitions from the \emph{Undiagnosed} state to the \emph{Diagnosed} state. Here,
the denominator is the number of persons in the \emph{Undiagnosed} state. Hence it holds
$\lambda' = \tfrac{p_1}{p_0 + p_1} \, \lambda_1.$ As $\lambda_0$ is the overall incidence, 
it holds 
$$\CDR = \frac{p_1 \, \lambda_1}{(p_0 + p_1) \lambda_0} = \wp_2 \, \DR_0.$$ 



\end{document}